\documentclass[
    ,final            
  ]
  {aipproc}
\usepackage{amsmath}
\layoutstyle{6x9}

\def\E{\textbf{E}\ }
\def\B{\textbf{B}\ }
\def\F{\textbf{F}\ }
\def\J{\textbf{J}\ }
\def\W{\textbf{\emph{w}}\ }
\def\d{\textbf{d}\ }
\def\RE{\textbf{R} } 
\def\R{R } 
\def\PR{\mathcal{R}} 

\begin{document}

\title{A gravitational explanation for quantum theory - non-time-orientable manifolds}

\classification{0240-k,0365.Ta,0420.Gz}
\keywords      {Topology, time-orientability, spin-half, quantum theory}

\author{Mark J Hadley}{
  address={Department of Physics, University of Warwick, Coventry
CV4~7AL, UK\\ email: Mark.Hadley@warwick.ac.uk}
}

\begin{abstract}
 Spacetime manifolds that are not time orientable play a key role in a gravitational explanation of quantum theory. Such manifolds allow topology change, but also have fascinating additional properties such as net charge from source-free equations and spin half transformation properties. It is shown how the logical structure of propositions and the probabilities of quantum theory arise from such acausal space times.
\end{abstract}

\maketitle

\subsection{Introduction}
Einstein's dream was to find a grand unified theory. For him that meant not only a unification of the forces of nature, but also the unification of particles and fields. He tried repeatedly to find field theories that included gravitation and also had particle-like solutions. It is interesting to note that some early attempts were dismissed because they did not have symmetric solutions of unequal masses and opposite charge corresponding to an electron and a proton - the only know elementary particles at the time.

Although the focus of his later work was on classical field theories, Einstein also held the view that a more fundamental realist theory would eventually explain quantum theory. Einstein probed the completeness of quantum theory with his EPR experiment, but it was Bell who used the same thought experiment to show that quantum theory was fundamentally incompatible with any local hidden variable theory. This ruled out any realist explanation compatible with causality.

To model elementary particles with a classical field theory requires solutions:
\begin{enumerate}
    \item With the right properties - such as mass, charge, spin etc
    \item Interactions between them - scattering, creation annihilation etc.
    \item The right behaviour - quantum theory
\end{enumerate}

Although Einstein considered the first requirement in detail, the other two were largely ignored. It may have seemed expedient to get solutions first and then examine interactions and mechanics afterwards. However, the second and third requirements place severe constraints upon acceptable solutions and in so doing offer potentially helpful clues.

\subsection{Interactions and topology change}

If particles are modeled as topological structures of space, then interactions such as annihilation of particles, particle-antiparticle creation and some scattering experiments will require the topology of a region of spacetime to change. However there are powerful theorems due to Geroch \cite{geroch} and Tipler \cite{tipler} that constrain topology change in classical general relativity. The possible counterexamples can be summarised as\cite{hadley98}:

\begin{enumerate}
\item Singularities
\item Closed timelike curves
\item A lack of time orientation
\end{enumerate}

The first is a breakdown of general relativity. Since General Relativity is expressed in terms of a 4D spacetime manifold, a singularity is not consistent with a description s spacetime as a manifold. Nevertheless some authors (eg Sorkin) \cite{sorkin97} have proposed a reformulation of general relativity to alow singularities where topology change takes place.

For closed time-like curves to appear in a region of space that was previously regular requires negative energy due to Tipler's theorem. While this cannot be ruled out absolutely with current theoretical knowledge it is unwelcome as a postulate.

The third option is that some timelines from the initial surface turn around and return through the same surface. This would be a failure of time-orientability. Further work on structures that lack time orientability have proved fascinating and fruitful.

Manifolds that do not admit a time orientation can be constructed in a number of ways. The Einstein Rosen bridge is the earliest example, although its non-orientability is often overlooked or misunderstood. The common description of a path through the bridge is that the traveler goes into a black hole and comes out of a white hole. In fact both ends of the wormhole structure are blackholes, but the traveler has his time direction reversed, making all blackholes look like white holes.

Wormholes can be constructed with any combination of time and space orientability\cite{visser}, and by a similar process monopole type structures of any orientation can be described\cite{diemer_hadley}. The M\"{o}bius strip is non-orientable and if the central circle is considered to be a space dimension, and the other direct a time direction, then the M\"{o}bius strip models a $1+1$ dimensional spacetime that is space orientable but not time orientable.

\subsection{Orientability and charge}
Manifolds that are not time orientable have some fascinating properties. They provide counterexamples to the  integral theorems linking the enclosed charge to the flux on a closed surface. When the theorems do not apply, it is possible to have the appearance of charge arising from the source free equations because there can be a net surface flux with zero enclosed charge. Sorkin \cite{sorkin} applied the idea using Stokes' theorem to a geon that was not space orientable and found that it could have net magnetic charge. However a 4-geon that is not time orientable does not have a consistently defined normal vector and the divergence theorem fails, allowing the appearance of net electric charge. A mathematical treatment in arbitrary dimensions is given in \cite{diemer_hadley}.

Working in three dimensions with the vector form of Maxwell's equations, the apparent source of charge is a net flow of flux into or out of a boundary region. A Sphere is the simplest example although all the results apply to an arbitrary compact two dimensional surface that encloses, or appears to enclose a volume of three space. The charge could be either magnetic or electric.
%

Considering first the magnetic charges $Q_m$:
\begin{equation}
Q_m = \oint_{S^2 = \partial V^3} \B.\hat{\textbf{n}}\  \rm{dS}= \int_{V^3} \textbf{Div}. \B \rm{dV} = 0\\
\end{equation}

which neatly relates the flux at a surface to the integrated charge density in the volume enclosed. It shows the apparent charge being due to the sources in the equations. The last step is a simple point by point application of Maxwell's equations. The first step is a global result that requires that the volume, \rm{dV}, is compact and orientable. The integrals also require a metric to be defined. In general the equations do apply and there are no Magnetic charges as a consequence of the vanishing divergence of the magnetic vector field.

A similar treatment for the electric charge gives:
\begin{eqnarray}
Q_e = \oint_{S^2 = \partial V^3} \E.\hat{\textbf{n}}\  \rm{dS} = \int_{V^3} \textbf{Div}. \E \rm{dV} = \int_{V^3} \rho \rm{dV}
\end{eqnarray}

which relates the electric flux at a surface to the integrated charge density in the volume enclosed. As in the case above the volume has to be compact and orientable. But the volume integral also requires a consistent definition of the electric field. However the electric field depends upon the direction of time, and reverses if time is reversed. So it is not necessarily possible to give a global definition of the charge throughout the volume - it will be impossible if the field is non-zero. Consequently, on a non time orientable spacetime, it is possible to have the appearance of electric charges without any sources.

%
The transformation properties of $\E$ are evident from the Lorentz invariant description using the Faraday tensor %
which shows that the electric field is the space.time component of the tensor and therefore changes its value depending on the direction of the time coordinate. In the absence of a global time direction, the electric field cannot be consistently defined even when F is well defined everywhere. By contrast the magnetic field is the space.space components of the field and does not depend upon the time direction (nor even the space direction). Using the Faraday tensor, Maxwell's equations take the simple form: $\d \F =0$ and $\d \star \F = \star \J$. The Faraday tensor is a two form and can be integrated using Stokes' theorem, which applies in any k-form, $\W$ and(k+1) dimensional volume, $V$:
\begin{equation}
   \int_{S= \partial V} \W = \int_V \d \W
\end{equation}

Stokes' theorem requires $V$ to be a compact oriented manifold. (interestingly it does not require a metric).

Magnetic charge can be defined in terms of the Faraday tensor:

\begin{equation}
Q_m = \int_{S= \partial V} \F  = \int_V \d \F = 0
\end{equation}

While a similar application to Stokes' theorem for the dual tensor gives:

\begin{equation}
Q_e = \int_{S= \partial V} \star \F =\int_V \d \star \F
\end{equation}
The RHS evaluates to zero in the absence of any sources ($\J =0$). Although the criteria for Stokes' theorem seem to be met for a manifold that is not time orientable, the star operator is defined in terms of an oriented space-time volume element so it is not well defined if time is not orientable and so the volume integral is not well defined.

The two treatments are equivalent, as are other approaches using vector densities. The conclusion is the same, that manifolds that are not time orientable can exhibit net electric charge from the source free equations, while the divergence free magnetic field still implies that there can be no net magnetic charge.

\subsection{Spin-half transformation properties}

The rotational properties of a non-time orientable manifold are intriguing. Intrinsic spin is a measure of how an object transforms under a rotation. Trying to apply the concept of a rotation to a manifold is not trivial. If a particle is modeled by an asymptotically flat manifold with non trivial topology (Wheeler used the term geon) or even non-trivial causal structure (a 4-geon), then we can define a rotation as any transformation with appropriate continuity properties, that matches the usual definition, $\RE(\theta)$, in the asymptotic region of the manifold.

\begin{eqnarray}
 & &\R(\theta )M \to M \\
 & &\R(\theta )\R(\phi )x = \R(\theta  + \phi )x \ \  \forall x \in M \\
 & &\R(0)x = x \ \  \forall x \in M \\
 & &\R(\theta )x \to {\RE } (\theta )x \ \  \text{as}\ |x| \to \infty
\end{eqnarray}

The axis of rotation has been omitted, a thorough treatment requires a definition that is manifestly applicable for the full rotation group, but a restriction to a single axis is adequate for our purposes. The definition above is very general allowing a very wide range for mappings to be counted as a rotation.

The definition above may suit mathematicians, but it is not appropriate for a physical rotation, such as when a neutron is rotated in the laboratory by a magnetic field. The distinguishing feature of a physical rotation is that it is parameterised by time. The object is rotated from zero degrees to $\theta$, as time passes from 0 to 1 say. So we refine our definition to a mapping of spacetime points: $\PR (\theta )(x,0) \to (\R(\theta )x,1)$, plus the conditions above.

%


The mathematical rotation defines a path through space, from the original point $x$ to the rotated point: $x \to \R(\theta )x,\ \ \gamma _\theta  (\lambda ) = \{ \R(\lambda \theta )x:\ \ x \in M,\lambda  \in [0,1] \}$\ %
%
%
Similarly, the physical rotation defines a world line starting at each point in spacetime: $(x,0) \to \PR(\theta)(x,0) = (\R(\theta )x,1), \ \ \chi _\theta  (t) = \{ (\R (t \theta) ,t)x:x \in M,t \in [0,1]\}$
%

But this construction defines a time direction throughout the manifold. It cannot apply to a spacetime that is not time orientable. A modification of the definition is required:

\begin{eqnarray}
  & & \PR (\theta )(x,0) \to (\R(\theta )x,\phi (\theta )(x)) \\
  & & \phi (\theta )(x) \to 1\ \ \text{as}\ |x| \to \infty
\end{eqnarray}

Where $\phi$ is zero a time direction is not defined by the rotation. The function $\phi$ must be zero over a two dimensional subspace if the manifold is not time orientable. The rotations have two categories of points that do not move:
\begin{description}
  \item[Fixed Points] $(x,0) \to (x,t)$ for example the points on the axis of the rotation.
  \item[Exempt Points] $(x,0) \to (x,0)$ in a sense these points do not participate in the rotation.
\end{description}

Consequently, a physical rotation that is 360 degrees in the asymptotic region will leave some exempt points in the manifold, and it will not be possible to find a 360 degree rotation that is the same as the identity, zero degree rotation. The topology of the rotation group is such that it is possible to find a 720 degree rotation that is an identity.
%
%

The idea has been known for centuries and is modeled by Wheeler's cube in a cube \cite{MTW}, Feynman's scissors trick \cite{feynman_weinberg} and Hartung's tethered rocks \cite{hartung}. A computer animation can be seen on the web \cite{spinhalf}.  What all the demonstrations have in common is a set of fixed points that are not rotated, but all lack any explanation of how parts of an elementary particle could be anchored in free space. This model of particles as 4-geons provides an explanation and naturally describes particles with spin half.

It is notable that the requirement for non time orientable manifolds came from the desire to model elementary particles and quantum phenomena. The existence of charge and spin half arise as a natural consequence.

\subsection{The Essence of Quantum Theory}

Quantum theory has many strange features, but most are not absolutely unique to quantum theory. For example the uncertainty relationship between position and momentum is common to classical waves, or the wave equations themselves. While some, apparently critical, features are largely irrelevant; such as the inability to make simultaneous measurements of some observables - if this were possible  (even for non-commuting observables) it would not change quantum theory.

A fundamental difference between quantum and classical physics can be found in the logical structure of the propositions. Classical physics satisfies Boolean logic while quantum theory is non-Boolean. The distributive law does not hold, instead the propositions form an orthomodular lattice \cite {beltrametti_cassinelli}. This leads to the requirement to represent probabilities as subspaces of a [complex] vector space rather than as measures on a volume space. It is this fundamental distinction that leads to the dynamical and statistical features of quantum theory.

 Although the relationship between incompatible observables has a non-Boolean characteristic, in any one experiment the propositions satisfy the normal Boolean logic. This is generally expressed by saying that quantum theory is {\em context dependent} - within any single context, the probabilities can be expressed in the normal way.

 A consequence of the logical structure of quantum theory and the way probabilities are represented is an entirely new concept of probability. In all of classical physics probability can be ascribed to our ignorance of some variables (typically the precise initial conditions). In quantum theory this interpretation is not possible, it cannot be described in terms of local hidden variables. As shown by the violations of Bell's inequalities.

The logical structure and the new way to represent probabilities leads inevitably, and uniquely,  to quantum theory and quantum field theory. The only known way to represnt probabilities of such non-Boolean logic is to use spaces and subspaces of a Hilbert space \cite{beltrametti_cassinelli}. The continuity and symmetry of spacetime then give rise to the usual equations of quantum theory as is shown by \cite{ballentine} for the non-relativistic case, and by \cite{weinberg} for the relativistic case..

A formal proof that acausal spacetimes lead to the logical structure of quantum theory is given in the paper by Hadley \cite{hadley97}. The proof is formal and can be largely replaced by the statement that an acausal spacetime is context dependent. This can been seen simply in two ways:

With Closed timelike curves (CTCs) or a failure of time orientability, it is not in general possible to set up boundary conditions on an {\em initial} surface without some knowledge of {\em future} conditions. Future experiments can set extra boundary conditions that are not redundant. Simple models with Billiards such as those described by Carlini and Novikov \cite{carlini} provide a good illustration. The standing waves on a string provide a helpful analogy; boundary conditions need to be specified at both ends before the wave can calculated.

An alternative way to look at the context dependence, with the same example, would be to consider the shape of the standing wave as a combination of forward and backward moving waves. A change of time direction at a future experiment would {\em send a signal back} to the start of the experiment. This can be seen as a realisation of the Cramer's transactional interpretation of quantum theory \cite{cramer}.

%
%

\subsection{Conclusion}

The results described above apply to any geometric theory of space and time that allows non trivial topology. General Relativity is the established theory of spacetime and satisfies the criteria, but most variations of general relativity would give the same result.
At least in principle, General Relativity with non-trivial causal structure could explain quantum theory and much more besides. It may be the unified theory that Einstein sought for so long.





\bibliographystyle{aipproc}   




\end{document}